\documentclass[a4paper,11pt]{article}
\usepackage{jcappub} 
\usepackage[utf8]{inputenc}
\usepackage{amstext}
\usepackage{amssymb}
\usepackage{hyperref}
\hypersetup{
    colorlinks=true,
    linkcolor=blue,
    filecolor=magenta,      
    urlcolor=cyan,
}
 
\usepackage[T1]{fontenc} 

\usepackage[utf8]{inputenc}
\usepackage{graphicx}
\usepackage{float}
\usepackage{subfigure}
\usepackage{color}
\usepackage{xcolor}
\usepackage[normalem]{ulem}

\def\beq{\begin{equation}}\def\eeq{\end{equation}}
\def\bea{\begin{eqnarray}}\def\eea{\end{eqnarray}}

\newfont{\cursive}{pzcmi at 9pt}

\newcommand{\gn}[2]{#2}

\title{Pearson cross-correlation in the first four black hole binary mergers}

\author[a]{Paolo Marcoccia}
\author[b]{\!\!, Felicia Fredriksson}
\author[a, c]{\!\!, Alex B. Nielsen}
\author[a]{\!\!, Germano Nardini}

\affiliation[a]{Department of Mathematics and Physics, University of Stavanger, NO-4036 Stavanger, Norway}
\affiliation[b]{Department of Physics and Astronomy, Uppsala University, Box 516, SE-75120 Uppsala, Sweden}
\affiliation[c]{Max Planck Institute for Gravitational Physics, 38 Callin Street, D-30167 Hanover, Germany}

\emailAdd{paolo.marcoccia@uis.no}
\emailAdd{felicia.fredriksson.7699@student.uu.se}
\emailAdd{alex.b.nielsen@uis.no}
\emailAdd{germano.nardini@uis.no}

\date{August 2020}

\abstract{
We adopt the Pearson cross-correlation measure to analyze the LIGO Hanford and LIGO Livingston detector data streams around the
events GW150914, GW151012, GW151226 and GW170104. We find that the Pearson cross-correlation method is sensitive to these signals, with correlations peaking when the black hole binaries reconstructed by the LIGO Scientific and Virgo Collaborations, are merging. We compare the obtained cross-correlations with the statistical correlation fluctuations arising in  simulated Gaussian noise data and in LIGO data at times when no event is claimed.
Our results for the significance of the observed cross-correlations are broadly consistent with those announced by the LIGO Scientific and Virgo Collaborations based on matched-filter analysis. In the same data, if we subtract the maximum likelihood waveforms corresponding to the announced signals, no residual cross-correlations persists at a statistically significant level.}

\begin{document}
\maketitle
\flushbottom

\section{Introduction}

The detection of gravitational waves (GW) by LIGO was a major milestone in the history of astronomy~\cite{Abbott:2016blz}. Achieving the necessary strain sensitivity of the instruments was a memorable technological accomplishment \cite{TheLIGOScientific:2016agk}, while determining the required waveforms was an astonishing success of the physics community~\cite{Brugmann366}. Despite these great achievements, the detection of the signals remains a challenge: the signals are still at low signal to noise ratios and must be extracted from the data using advanced statistical techniques and signal processing.

The most sensitive gravitational wave data searches rely on matched-filtering techniques~\cite{Allen:2005fk, Usman:2015kfa, Sachdev:2019vvd}. These are based on comparing the data with a class of signals determined in a specified theory. Such techniques are robust and very sensitive if the source waveform is accurately predicted. For signals with an uncertain modelling, more model-independent, although less sensitive, searches are necessary~\cite{Klimenko:2015ypf, Lynch:2015yin, LIGOScientific:2019fpa,Salemi:2019uea, Tsang:2019zra, Edelman:2020aqj}. The coherent wave approach is particularly suitable in these cases. Its flexibility enables it to identify signals that deviate from model expectations. There are still some intrinsic assumptions in the coherent wave method (such as the waves being plane waves travelling at the speed of light with the polarisation behaviour of general relativity (GR) waves), so that more agnostic techniques are required to rule out or discover (unexpected) signals not fulfilling these assumptions. Searching for statistically significant correlations in the data of noise-uncorrelated detectors, with as few as possible biases on the signal structure, is poorly efficient in terms of reconstruction power and computing resources but is an  option that still deserves some attention. Moreover, the same searches applied to the data streams where the reconstructed signal is subtracted, are a powerful tool to quantify the quality of the reconstructions and the possibility of incoherent detector noise.

Matched-filtering searches typically assume Einstein's theory of general relativity. This theory is the standard paradigm for gravity, having passed all tests in the Solar System with flying colours. Gravitational wave observations can be used to test this theory in different systems. Neutron star binary systems have been observed via gravitational waves and other astronomical messengers. These observations are already raising questions about our theoretical understanding of neutron star populations~\cite{Monitor:2017mdv, Abbott:2020uma}. On the other hand, black hole binaries, triplets and other ``dark'' systems can be currently discovered only by means of gravitational wave observations. Robust confidence in the analysis of gravitational wave data is thus of key importance for the exact interpretation of the observations. This requires comprehending many aspects, for instance how deviations from expected models might manifest themselves in the data~\cite{Barausse:2014tra}, and how to be sensitive to the unexpected.

Monitoring the correlation in the data collected by independent interferometers constitutes an interesting method to test the unexpected. 
Some publications have explored this direction analyzing the LIGO data streams with known events~\cite{Liu:2016kib, Creswell:2017rbh, Liu:2018dgm, Nielsen:2018bhc, Jackson:2019xbq, Maroju:2019ymy}. 
It turns out that several subtleties jeopardize the reliability of such a test \cite{LIGOScientific:2019hgc}. For instance, after supposedly subtracting the GW150914, GW151012 and GW151226 signals from the corresponding data streams, Refs.~\cite{Liu:2016kib, Creswell:2017rbh, Jackson:2019xbq} find some anomalies that could potentially be ascribed to a sizable mismatch between the detected signals and the general-relativity waveforms that best fit these signals. However Ref.~\cite{Nielsen:2018bhc} concludes that this mismatch for GW150914 is not statistically significant once one takes care of subtleties, such as whitening versus notching or discrepancies in the best fit waveforms. Still, some debate remains over the exact size of the statistical significance~\cite{Jackson:2019xbq, Maroju:2019ymy}.

The question whether the anomalies claimed in Ref.~\cite{Creswell:2017rbh} for GW151226 and GW170104 are actually caused by similar subtleties, has not been fully addressed yet. In the present paper we tackle this question, and  for completeness we run our tests also on GW151012, a black hole merger detected in the first observing run, that was initially classified as marginal but has now been promoted to a confident detection by several groups \cite{Nitz:2018imz, LIGOScientific:2018mvr, Venumadhav:2019tad}.\footnote{We restrict our analysis to the events GW150914, GW151012, GW151226 and GW170104  as their LIGO data can be easily handled by means of the Gravitational Wave Open Science Centre (GWOSC) toolbox~\cite{Vallisneri:2014vxa}. The GWOSC  webpage indeed provides several numerical tools already tuned for the analyses of these four events.
There is no conceptual obstacle in analyzing other events in the way we present here.}
We believe that our results help to clarify several aspects that can guide the community in future implementations of  correlation-based searches for new physics, in the presence and in the lack of a signal model.

This  manuscript  is  organized  as  follows.   Section~\ref{sec:int_meth} details how we preprocess the data and apply the Pearson cross-correlation measure. Section~\ref{sec:results} contains the results and comments on them.  Finally, Section~\ref{sec:conclusions} is devoted to our main conclusions and outlook.

\section{Methodology}
\label{sec:int_meth}

Our methodology follows closely the approach presented in Ref.~\cite{Nielsen:2018bhc} which itself is designed to scrutinize the claimed anomalous correlations in the GW150914 data from Refs.~\cite{Liu:2016kib, Creswell:2017rbh}. 
The approach in Ref.~\cite{Nielsen:2018bhc} is thus tuned to work on the features of the GW150914 signal. To extend it beyond GW150914, we need to decide some criteria aimed at generalising this procedure to other events.

In brief, the procedure we adopt is the following: we take the GWOSC calibrated gravitational wave strain data of LIGO Hanford and LIGO Livingston around the events GW150914, GW151012, GW151226 and GW170104; we whiten and bandpass the data of each event and each detector individually; we calculate the Pearson cross-correlation between the data streams of the detectors before and after subtracting best-fit general relativity waveforms; and we check the statistical significance of the obtained correlations with the statistical fluctuations  arising in pure noise. We now describe the choices of these steps and their reasons.

We choose here to whiten the data, rather than notch them with a list of frequency lines.
This simplifies our analysis, providing a unique criterion applicable to every event without requiring a complete list of individual lines to be notched. Such a list, varying from event to event, does not exist in general and would need to be generated by hand with some judgement
(we note that Ref.~\cite{Creswell:2017rbh} also uses whitening for their analysis of GW151226 and GW170104). This generalization of the analysis leads to a change of spectrum of both noise and templates.
In particular, the whitening reduces the contribution of low
frequencies to the residuals and cross-correlations. Whitening the data also more closely follows the practice in the wider gravitational wave 
literature; see e.g.~Ref.~\cite{LIGOScientific:2019hgc}. We refer to Ref.~\cite{Nielsen:2018bhc} for details on the whitening process we employ.

We apply a bandpass to the data filtering frequency ranges that varies for each event.  The chosen ranges loosely follow those considered in the GWOSC webpage~\cite{Vallisneri:2014vxa} which itself does not necessarily implement the values used in the published LIGO-Virgo Collaborations (LVC) analyses, but provides a convenient guide for our investigation.
The high-pass frequency adopted in GWOSC is $43$\,Hz but the data at  frequencies $35-60$\,Hz are usually highly dominated by seismic noise. We then round the high pass frequency to 50\,Hz for all events but the first one, for which we keep the choice adopted in Ref.~\cite{Nielsen:2018bhc} to facilitate the comparison between our findings and those previously obtained.
For the low-pass frequencies, we typically use \gn{larger}{lower} frequencies than those chosen by GWOSC, which are 300\,Hz for GW150914, 400\,Hz for GW151012 and 800\,Hz for GW151226 and GW170104.\footnote{These values correspond to the option `fband' in the json files of the GWOSC webpage~\cite{Vallisneri:2014vxa}.} 
We in fact take a low-pass frequency not smaller than 
$200-300$\,Hz for these frequencies in order to be free of seismic and thermal noise, and sufficiently below the violin frequency disturbances at around 500\,Hz.\footnote{We ran some tests on power spectral density at about the time of the GW151226 event. For a low pass of 480\,Hz or larger, the data at high frequencies are still very noisy due to contamination at the violin frequencies.}

It is of course essential to not cut out the dominant frequencies associated with the binary coalescence. Indeed a cross-correlation analysis, as opposed to an extended matched-filter, appears to be more sensitive to the signals around their peak of strain. For a binary merger it is hence weakly impacted by the low-frequency strain at the start of the inspiral phase whereas it is dominated by the higher-frequency strain around the coalescence stage. Table~\ref{tab:table1} shows the specific values of the low and high bandpasses we implement.\footnote{As a consistency check, for the GW151012 event we ran our analysis with different choices of high- and low-pass frequencies. For variations of about $\pm5$\,Hz, the findings do not qualitatively change.}
These values are in part determined by the inferred properties of the signal and thus further investigations with truly blind searches would be required to test the capabilities of the Pearson correlation search method for unknown signals.

To construct the data sets of the residuals, we subtract best-fit general relativity waveforms. As shown in Ref.~\cite{Nielsen:2018bhc}, for GW150914 the maximum likelihood waveform produces cleaner residual data than the numerical relativity waveform released on GWOSC. In our analysis here we subtract maximum likelihood waveforms, specifically the one provided in Ref.~\cite{De:2018zrk} for GW170104 and those available in Ref.~\cite{Biwer:2018osg} for the other three events.  

To measure the cross-correlations between the Hanford and Livingston data sets, we apply a Pearson cross-correlation measurement. The Pearson cross-correlation is given by
\begin{equation}
    R(\tau, t, \omega) = \int_{t}^{t+\omega}{\frac{H(t^{\prime} + \tau)}{\sigma_H} \frac{L(t^{\prime})}{\sigma_L}dt^{\prime}}~,
\label{eq:corrR}
\end{equation} 
where $H(t)$ and $L(t)$ are respectively the data in Hanford and Livingston (with or without the signal subtracted) at the time $t$ set in Hanford, ${\sigma_H}$ and ${\sigma_L}$ are the standard deviations of the Hanford and Livingston data respectively, and $\omega$ is the time window in which the correlation is measured. The quantity $\tau$ is the time lag between the signal arrival times in Hanford and Livingston (positive if the signal reaches Livingston first). Unfortunately, in most of the realistic cases,  $\tau$ comes with an important uncertainty $\Delta\tau$ that Eq.~\eqref{eq:corrR} does not take into account.  
We thus need to modify the measure $R$, and  define the (improved) cross-correlation measure
\begin{equation}
\label{eq:corrC}
    C(\tau, \Delta \tau, t, \omega) =  \max_{\bar \tau} R(\bar\tau,t,\omega) \: : \: |\bar \tau-\tau|\le \Delta \tau~.
\end{equation} 

Due to computational limitations, we have not performed a complete multivariate analysis on $t$ and $\tau$. We instead prefer to analyse the $t$ dependence of $C$ along a strip $\tau_p \pm \Delta \tau_p$ --- with $\tau_p$ and $\Delta \tau_p$ taken from Refs.~\cite{Creswell:2017rbh,   TheLIGOScientific:2016pea}, and identify the time $t=t_p$ at which $|C(\tau_p, \Delta \tau_p, t, \omega)|$ is maximal. This facilitates comparison with existing results in the literature and makes manifest the impact of some subtleties such as different bandpasses, whitening and subtracted waveforms. On the other hand, the data preprocessing in Refs.~\cite{Creswell:2017rbh, Nielsen:2018bhc} and in ours differs in several aspects, so that it is not guaranteed that taking $\tau = \tau_p$ is a valid assumption. For this reason, we perform some consistency checks, and test a posteriori that at $t=t_p$ no higher cross-correlations arise when moving $\tau$ outside the strip $\tau_p\pm \Delta_p$.

To identify $t_p$, it is important to evaluate $C$ in a $t$ interval including the merger time around which the maximal cross-correlation is expected in the data before subtracting the reconstructed signal. In principle the length of this interval should mildly affect the findings, however an appropriate interval helps avoid statistical artifacts and include relevant signal data.

For concreteness we take $t\in [t_r - 0.1\, \textrm{s},t_r +0.1\, \textrm{s}]$, with $t_r$ being the arrival time (rounded at the first decimal digit) of the maximal signal strain in Hanford as estimated in the GWOSC analysis~\cite{Vallisneri:2014vxa}. 
For $\tau$ and $\Delta \tau$ we adopt the values obtained in Refs.~\cite{Creswell:2017rbh, TheLIGOScientific:2016pea}), while for $\omega$ we consider the time window $\omega= 40$\,ms which corresponds to four cycles of a gravitational wave of 100\,Hz. As a consistency check we allow $\tau$  to move from $-10$ to $10$\,ms. Such an analysis is performed for both the data sets with the signals and those with only the residuals.
Table~\ref{tab:table1} summarizes the inputs we adopt in our analysis.

\begin{table}
\centering
\begin{tabular}{|c | c | c | c | c | c | c |}
 \hline
 \textbf{Event} & \textbf{Bandpass} & \textbf{H-L time} & \textbf{Reference} & \textbf{Window} \\
 \textbf{Name} & \textbf{range (Hz)} &   \textbf{$\tau_p \pm \Delta \tau_p$ (ms)} & \textbf{time $t_r$ (s)} & \textbf{$\omega$ (ms)}\\
 \hline
 GW150914 & 35-350 &  ~6.9 $\pm$ 0.5  & 1126259462.4 & 0.4\\  
 GW151012 & 50-230 &  -0.6 $\pm$ 0.6 & 1128678900.4  & 0.4\\
 GW151226 & 50-460 &  ~1.1 $\pm$ 0.3 & 1135136350.6   & 0.4\\
 GW170104 & 50-230 &  -3.0 $\pm$ 0.5 & 1167559936.6 & 0.4\\
 \hline
\end{tabular}%
\caption{The values of the input parameters for the analysis of each event.}
\label{tab:table1}
\end{table}

Lastly, we  analyze the cross-correlation among the detectors for pure noise by estimating the backgrounds, following the procedure described in Ref.~\cite{Nielsen:2018bhc}.
However, our analysis differs from Ref.~\cite{Nielsen:2018bhc} in some technical points.
While in Ref.~\cite{Nielsen:2018bhc} the residual data was used for background times, we preferred to use the original data strain in a time interval far away from the coalescence, where no event was declared.
In particular, the time interval was chosen to start 12 minutes away from the coalescence time.

We generate several pairs of data sets, each pair consists of a set of mock data of simulated pure Gaussian noise and a set of LIGO data at times away from any claimed event. These pairs are then treated in exactly the same way as the event data we consider. We thus whiten and bandpass them, and run the $C$  estimator over them according to Table~\ref{tab:table1}. This tells us how often a given cross-correlation value arises as a statistical fluctuation in background noise, and furnishes
a probability estimate for the maximal correlations we find in our analysis to occur by chance in pure noise.

The code with the implementation of our analysis is public~\cite{pipeline}. We refer the reader to this for further details on the preprocessing of the data and their analysis.

\section{Results}
\label{sec:results}

In this section we discuss our main results. We start by presenting the $t$ dependence of $C$ in the strip $\tau_p \pm \Delta \tau_p$,
and then show the consistency check previously described. At the end we quantify the statistical significance of the identified cross-correlations before and after subtracting the maximum-likelihood waveforms of the reconstructed signals.

\subsection{Cross-correlation: $t$ dependence} 
\label{sec:t-dep}

\begin{figure}[tb]
  \centering
  \includegraphics[width=\columnwidth]{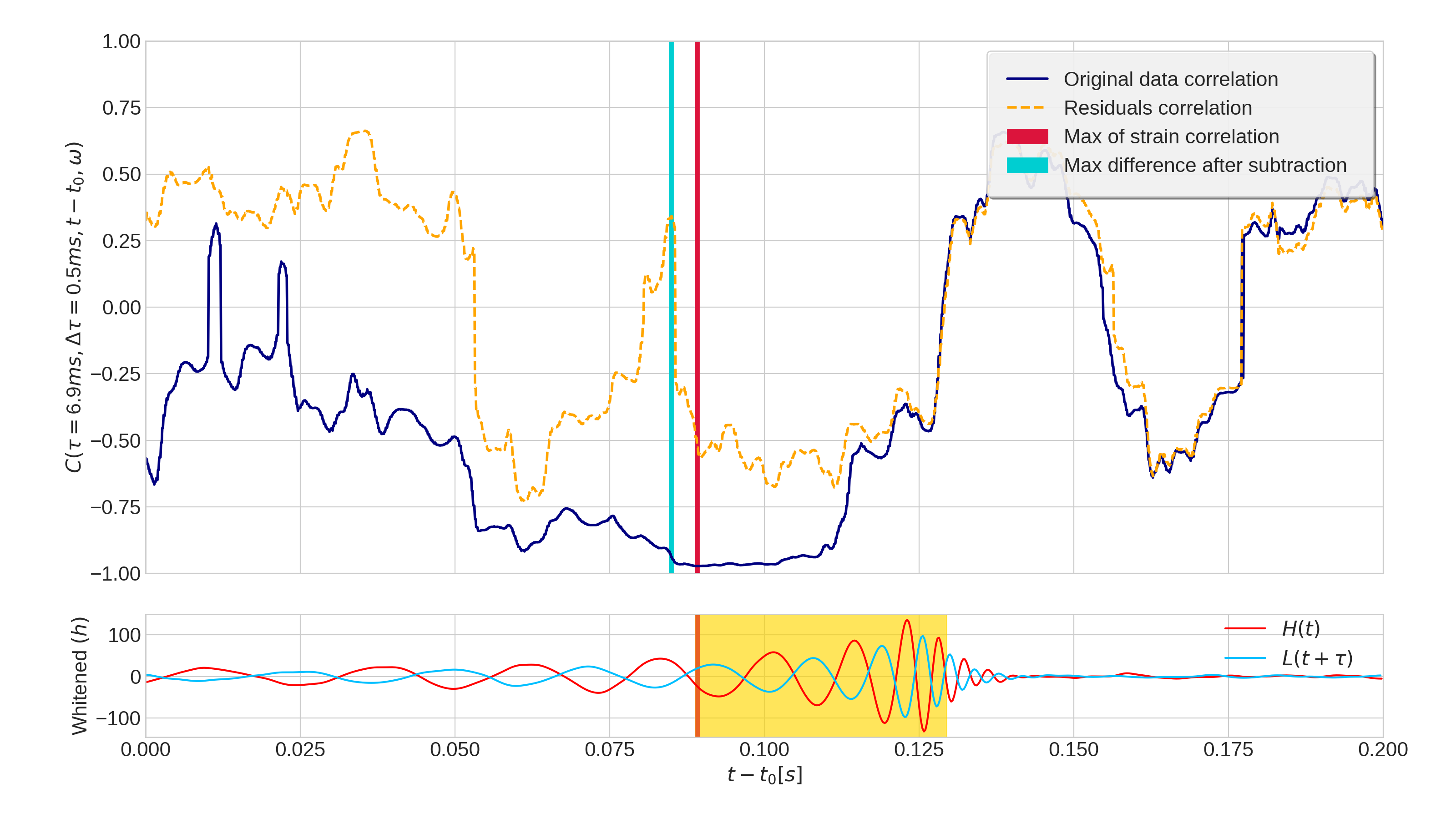}
\caption{Top panel shows the correlations between Hanford and Livingston detector strain data for $200$\,ms of data around GW150914. The bottom panel shows the whitened model waveforms that are subtracted from the data to produce the residuals, where the time interval highlighted in yellow represents the part of the waveforms that was integrated in the estimation of the correlation at the time of maximum correlation $t_p$. The Livingston waveform is shifted in time relative to the Hanford waveform in order to highlight the anti-correlation between the two strains.}. 
\label{fig:150914corr}
\end{figure}

\begin{figure}[h]
  \centering
   \includegraphics[width=\columnwidth]{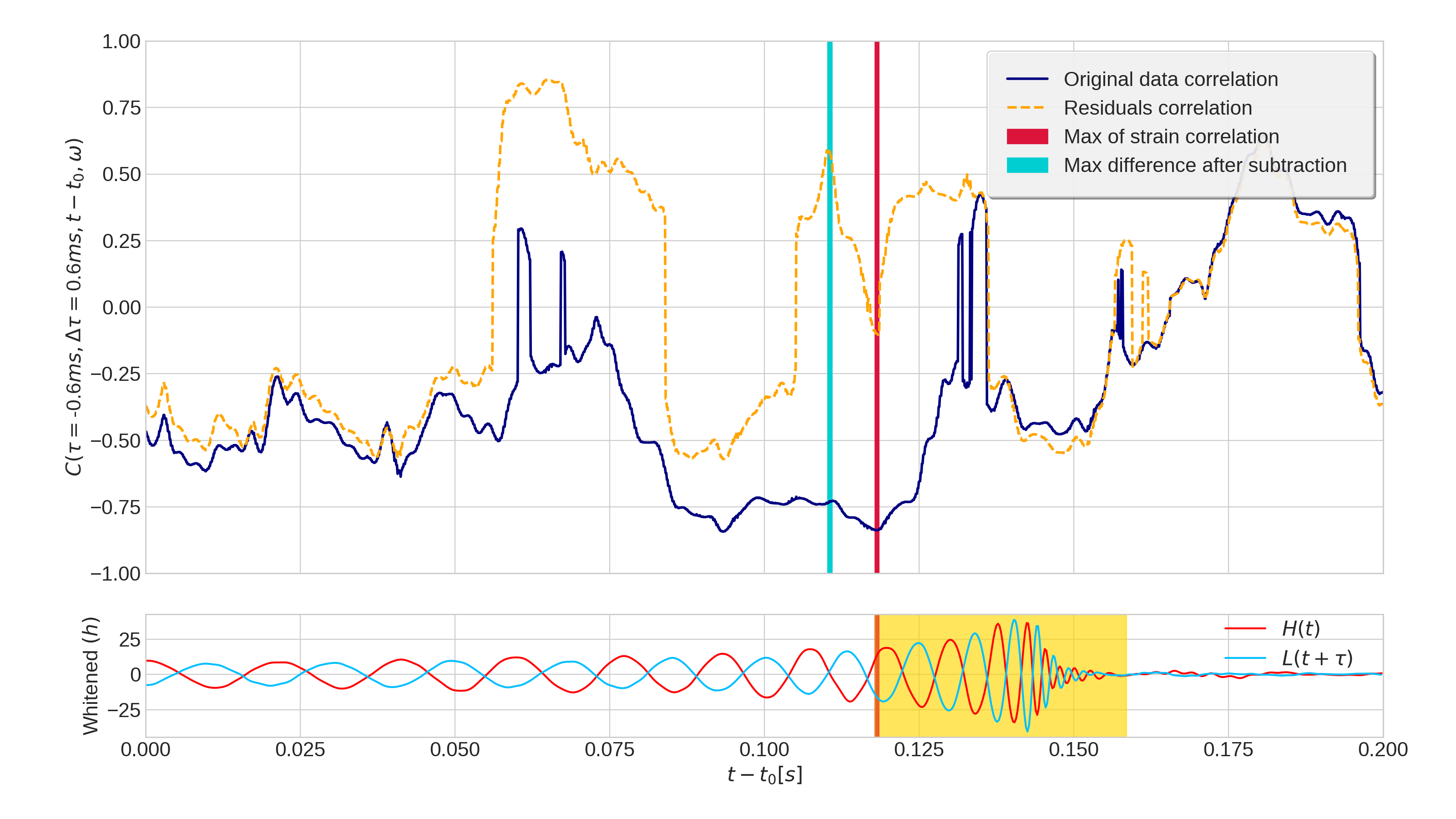}
\caption{As Fig.~\ref{fig:150914corr}, but for the GW151012 event.}
\label{fig:151012corr}
\end{figure}

\begin{figure}[t]
  \centering
  $~$\\[1cm]
    \includegraphics[width=\columnwidth]{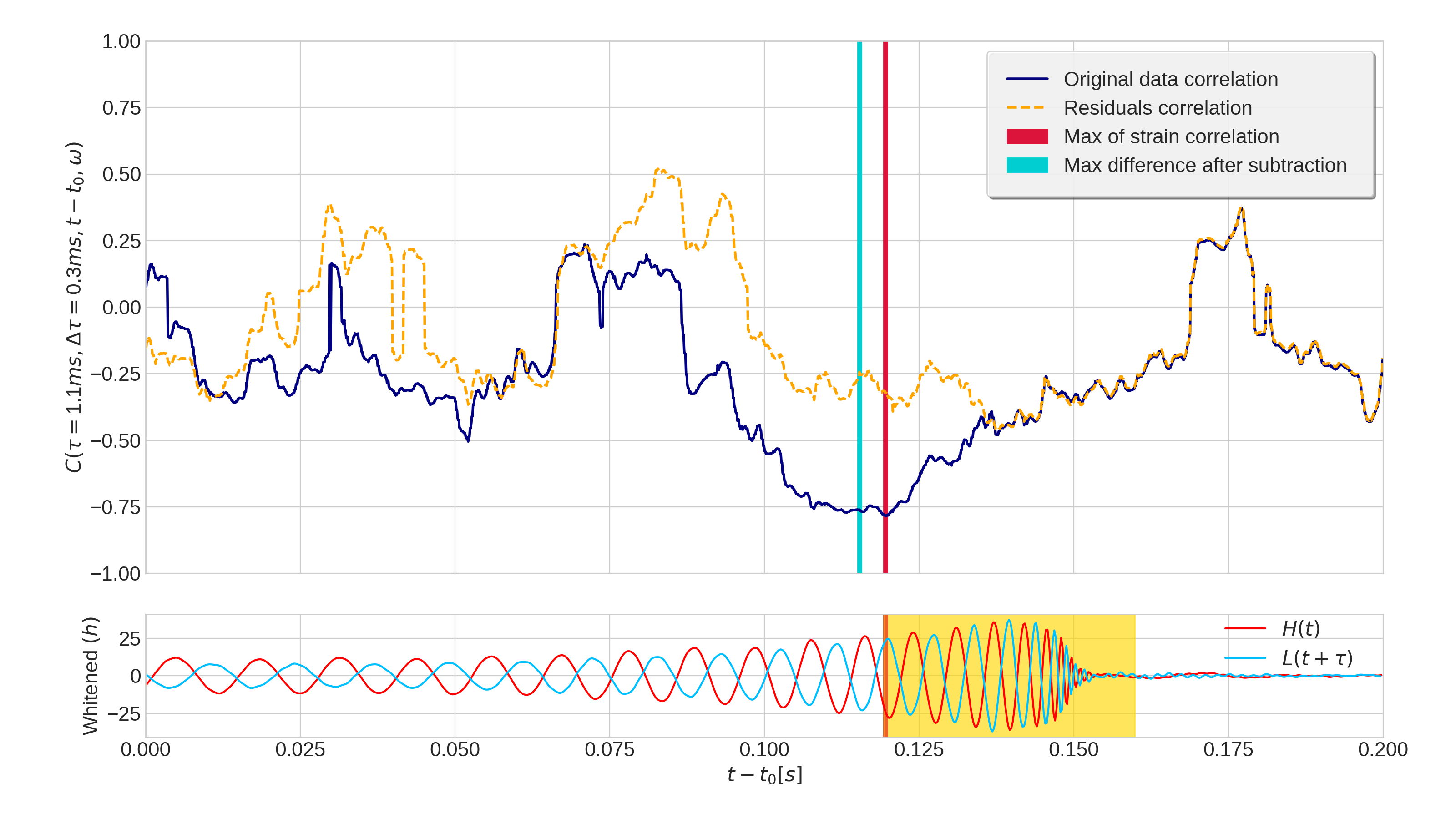}
\caption{As Fig.~\ref{fig:150914corr}, but for the  GW151226 event.}
\label{fig:151226corr}
\end{figure}

\begin{figure}[h]
  \centering
    \includegraphics[width=\columnwidth]{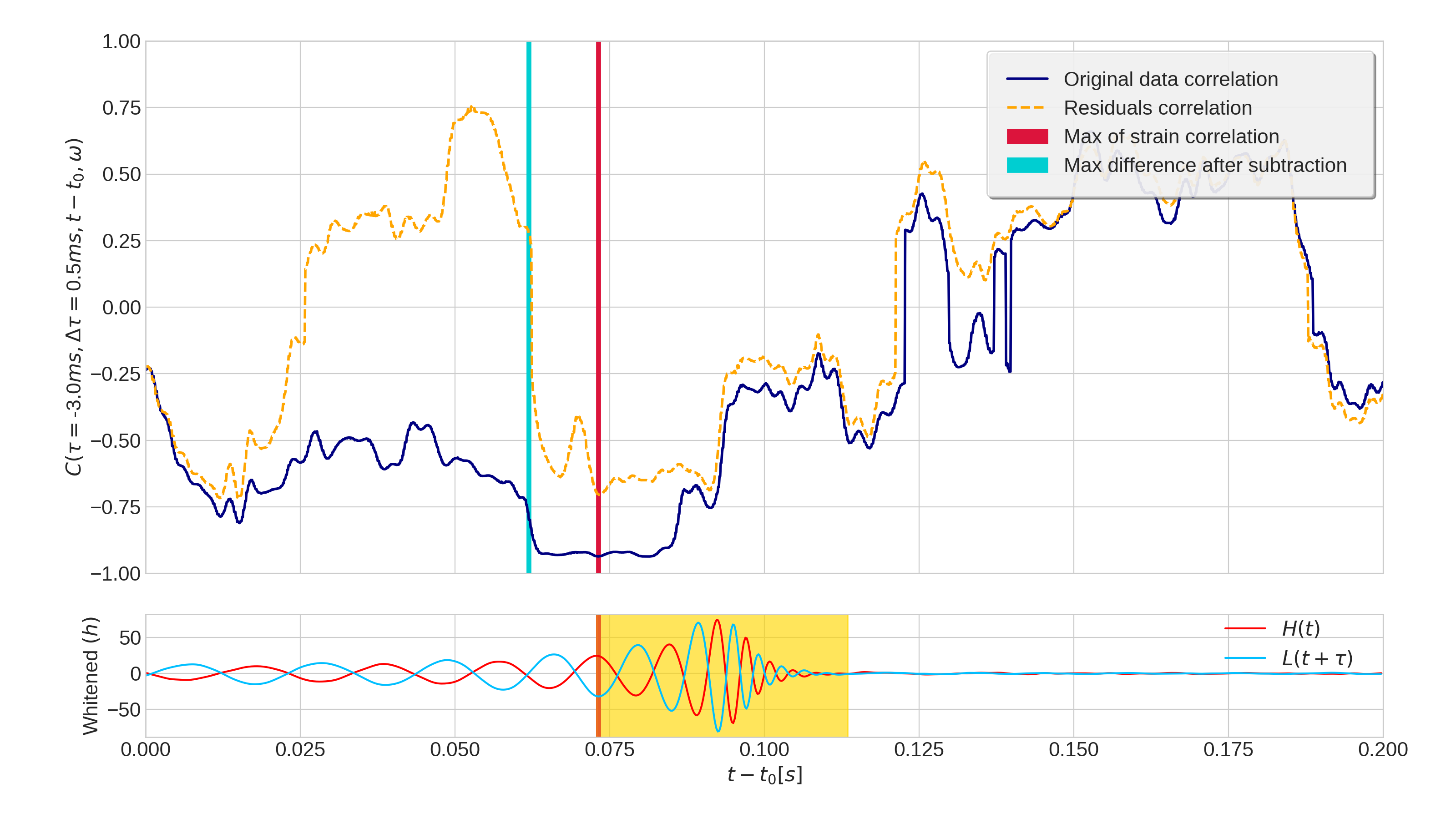}
\caption{As Fig.~\ref{fig:150914corr}, but for the GW170104 event.}
\label{fig:170104corr}
\end{figure}

We run the cross-correlation measure $C$ over the time series around the GW150910, GW151012, 
GW151226 and GW170104 LIGO data, preprocessed as explained in Section~\ref{sec:int_meth}.  The resulting $t$ dependence is plotted in the upper panels of Figs.~\ref{fig:150914corr}--\ref{fig:170104corr}, where the time variable is shifted by $t_0 \equiv t_r - 0.1$ for clarity. In each figure, each one dealing with a different event, the blue solid and dashed orange lines represent the $t$ evolution of $C$ in the data, respectively before and after subtracting the maximum likelihood waveforms of the reconstructed signals. The red vertical line marks $t_p$, the maximal (in absolute value) correlation time found in the data before the waveform subtraction. The green vertical line instead highlights $t_d$, the time at which subtracting the waveform signal leads to the maximal difference in the correlation between the original data and the residuals.
The waveform signals that are subtracted are depicted in the lower panels of the figures. They are shown after being whitened, bandpassed and, for the waveform measured in Livingston, shifted by the value of $\tau$ reported in Table~\ref{tab:table1}. In each of these lower panels, the yellow region highlights the time frame $\omega = 40$ms for the peak correlation value of $C(t_p, \tau_p, \Delta \tau_p, \omega)$ for the considered event.  By construction, such a time frame contains the data that the LIGO Hanford and LIGO Livingston detectors collect at the times $t \in [t_p,t_p+\omega]$ and $t \in [t_p+\tau_p,t_p+\omega+\tau_p]$, respectively.

For a signal that slowly increases in amplitude and then falls away rapidly, such as a gravitational wave inspiral, the choice of a forward integration will lead to a shift in time of the location of the maximum of the correlation integral compared to the location of the maximum amplitude of the signal. This may appear unusual in Figs.~\ref{fig:150914corr}--\ref{fig:170104corr}, because the maximum correlation is found at a GPS time that is shifted to the left of the real coalescence time, by a quantity approximately equal to the used integration window. Both the residual correlation, and the difference between the strain correlation and the residual correlation, fall off before the reference coalescence time $t_r$, since with a forward integration large parts of the integrated strain beyond that point come from regions where the signal template is rapidly diminished. In spite of this issue, we choose here to keep a forward integration for our correlation formula \ref{eq:corrR}, in order to enable easy comparison of our results with existing ones from the literature \cite{Liu:2016kib, Creswell:2017rbh, Jackson:2019xbq,Nielsen:2018bhc}.

With this choice we expect $t_p$ to be around $t_r - \omega$, i.e.~in the data segment preceding the signal's maximal strain (during the ringdown the amplitude of the signal is much weaker than in the late inspiral and merger phase). Our results fulfill this expectation (c.f.~$t_r$ and $t_p$ in Table~\ref{tab:table1} and Table \ref{tab:table2}, which quotes the key quantities calculated in this section).

Figure~\ref{fig:150914corr} shows the correlation versus detector time around the event GW150914. The cross-correlation behaviour of this event is widely discussed in the literature~\cite{Liu:2016kib,  Creswell:2017rbh, Nielsen:2018bhc, Biwer:2018osg, Jackson:2019xbq}. We present it here for ease of comparison with previous results. We notice that, after subtracting the reconstructed signal, the correlation in the time window $[t_p, t_p+\omega]$ is strongly reduced and the global peak of the correlation is no longer inside this window. The maximal variation between the cross-correlations before and after the subtraction is also outside of this time window, i.e.~$t_d < t_p$. This is not surprising if after the subtraction of the model waveform from the original data, what remains is a pure noise correlation between the two detectors. The residual correlation, assuming a perfect subtraction of the signal waveform, may vary randomly in magnitude, as well as in sign, independently from the subtracted signal.

We have to emphasise however, that even though noise correlations and signal correlations are not linearly additive, they are still additive. Having a really low value of correlation difference after subtraction at the time $t_*$, therefore suggests that most of the correlations at said time were given by statistical fluctuations of the noise, which are indeed not the ones we are interested in this kind of analysis. We will see in particular in the case of GW151012, where the low signal-to-noise ratio (SNR) of the incoming wave allow the noise correlations to be comparable with the ones generated by genuine gravitational waves effects, that the first peak of maximum correlation found can be discarded due to the low significance difference of the correlations before after the subtraction. We detail the  statistical compatibility of the residuals with the instrumental noise in Section \ref{sec:BkgStt}.

In the top panel of Fig.~\ref{fig:150914corr} after the merging time, there are still some slight differences between the original data correlation (in blue) and the residuals correlation (in orange).
Since the model waveform is zero after the ringdown, one might expect there should be no difference between the correlations of the original data of Fig.~\ref{fig:150914corr} and the residual data. However, the process of whitening applied both to the original data and the model waveform, introduces small difference in both of the considered data (see the variation after the ringdown part of the model waveform in the bottom panel), that once integrated over the window $\omega$ to estimate the correlations will result in the small mismatching observable among the two. This feature of the analysis, is present in all of the reported figures ~\ref{fig:150914corr}--\ref{fig:170104corr}.\\


The analogous plots for the event GW151012 are presented in Fig.~\ref{fig:151012corr}. GW151012 is the event with the lowest SNR among the ones we investigate. The figure shows that we still obtain relatively high values of $|C|$, although not as high as for GW150914. As we will see later in this section, the significance of these values is also not as high as the ones for GW150914. This is qualitatively consistent with the interpretation of a lower amplitude signal due to a smaller mass system at a greater distance.

At $0.05\,\textrm{s} \lesssim t-t_0 \lesssim 0.075\,\textrm{s}$ the data correlation occasionally becomes positive, even though we expect that the effects of the incoming gravitational waves should result in an anticorrelation among the detectors. It has to be emphasized though, that the correlations of residuals and reconstructed waveforms are not linearly additive, hence this effect arises for this event due to the low amplitude of the gravitational wave signal in the detectors, compared to the noise, at times away from the coalescence. At $t-t_0 \lesssim 0.010\,\textrm{s}$  the signal dominates the noise in the detectors. Hence, even though the correlation of the residuals flips sign, the correlation in the data before the signal subtraction remains negative until it lines up with the residual strain after the coalescence time. Eventually, after the signal waveform fades out, no sizeable correlation remains.\footnote{Our results do not confirm the late-time correlation found in Ref.~\cite{Salemi:2019uea}. However our results are restricted to time lags $\tau$ consistent with the main event and are thus blind to correlations at other time lags, that could be interpreted as originating from other parts of the sky.} 

Furthermore, at $t-t_0 \lesssim 0.010\,\textrm{s}$ a first maximum of strain correlation appears, however, as discussed previously this peak can be neglected because most of its anticorrelation is given by a statistical fluctuation of the noise. This fact may be further elucidated by observing the bottom panel of Fig.~\ref{fig:151012corr}, which clearly shows that in relation to the first peak we are not even including the merging part of the waveform in the correlation data. This event is the only one for which the correlations in the residual data are positive around the time of the signal merger, that is, in one forth of our events the residual exhibits a positive correlation, which seems statistically unremarkable to us.

Concerning GW151226 and GW170104 (see Fig.~\ref{fig:151226corr} and Fig.~\ref{fig:170104corr}), the $t$ dependence of $C$  essentially presents no key qualitative features not already noted above for GW150914 and GW151012. 
The signal in GW151226 has a reconstructed total mass much smaller than the other three. The signal thus exhibits the largest gravitational wave frequency at the coalescence time, about 447\,Hz~\cite{Vallisneri:2014vxa}, and the smallest relative strain around the coalescence time. The maximum of $C$ is then the smallest in absolute value. The amplitude of the signal in GW170104 is instead much higher. Its source involves two highly massive black holes (of approximately $30$ and $20$ solar masses) and is the next to farthest among the four considered sources.

For binaries with masses inferred for GW170104 at the reconstructed luminosity distance, theory predicts that the gravitational wave signal has a peak strain amplitude at a frequency around 190\,Hz, where the noise of the detectors is quite low. Thus the correlation of the data around this event peaks at a rather high value, almost as high as the one in GW150914.

The straight comparison between the values of the $C$ peak in different events can be misleading. For instance, let us consider the case of GW150914 and GW170104. As just observed, $C$ peaks at similar values in these two events. It is however known that the matched-filter SNR in GW150914 is almost twice as high as the matched-filter SNR in GW170104~\cite{LIGOScientific:2018mvr}, so it may seem that this observation is not compatible with the similar values of the correlation peaks. However, correlation values in and of themselves are not directly comparable without considering the different bandpass frequency ranges and length of data for which the correlations are calculated. We further investigate this aspect in Section~\ref{sec:BkgStt}.

For the four considered binary merger events, if one knows the time lag $\tau \pm \Delta \tau$ and has an indicative time data segment where to look in, $C(\tau, \Delta \tau, \omega, t)$ reaches large values in the proximity of the time frame where the inferred signal strength becomes high. In addition, running $C$ on the residuals constitutes a valuable check of the signal model. In the case of our four events, if one subtracts the signal reconstructed assuming the theory of general relativity, the correlation maximum (in absolute value) moves away from the time interval where $C$ peaks. This is not yet a complete test of the model but it is certainly encouraging. We come back to this point after performing the consistency checks on the values of $\tau$ we have just adopted.

\subsection{Cross-correlation: $\tau$ consistency check} 
\label{sec:tau-dep}

We have determined  $t_p$ by assuming $\tau=\tau_p$ and $\Delta\tau=\Delta \tau_p$, with $\tau_p$ and $\Delta \tau_p$ taken from previous studies~\cite{Creswell:2017rbh,TheLIGOScientific:2016pea}.   
Here we present the consistency checks. We test that $|C(\tau, 0, t_p, \omega)| < |C(\tau_p, \Delta \tau_p, t_p, \omega)|$ for $\tau \in [-10\,{\rm ms},\tau_p - \Delta \tau_p] \cup [\tau_p + \Delta \tau_p , 10\,{\rm ms}]$, that is, no higher cross-correlations are found at $t=t_p$ for a time lag $\tau$ away from the strip $\tau_p\pm \Delta \tau_p$.

\begin{figure}[h]
  \centering
    \includegraphics[width=\columnwidth]{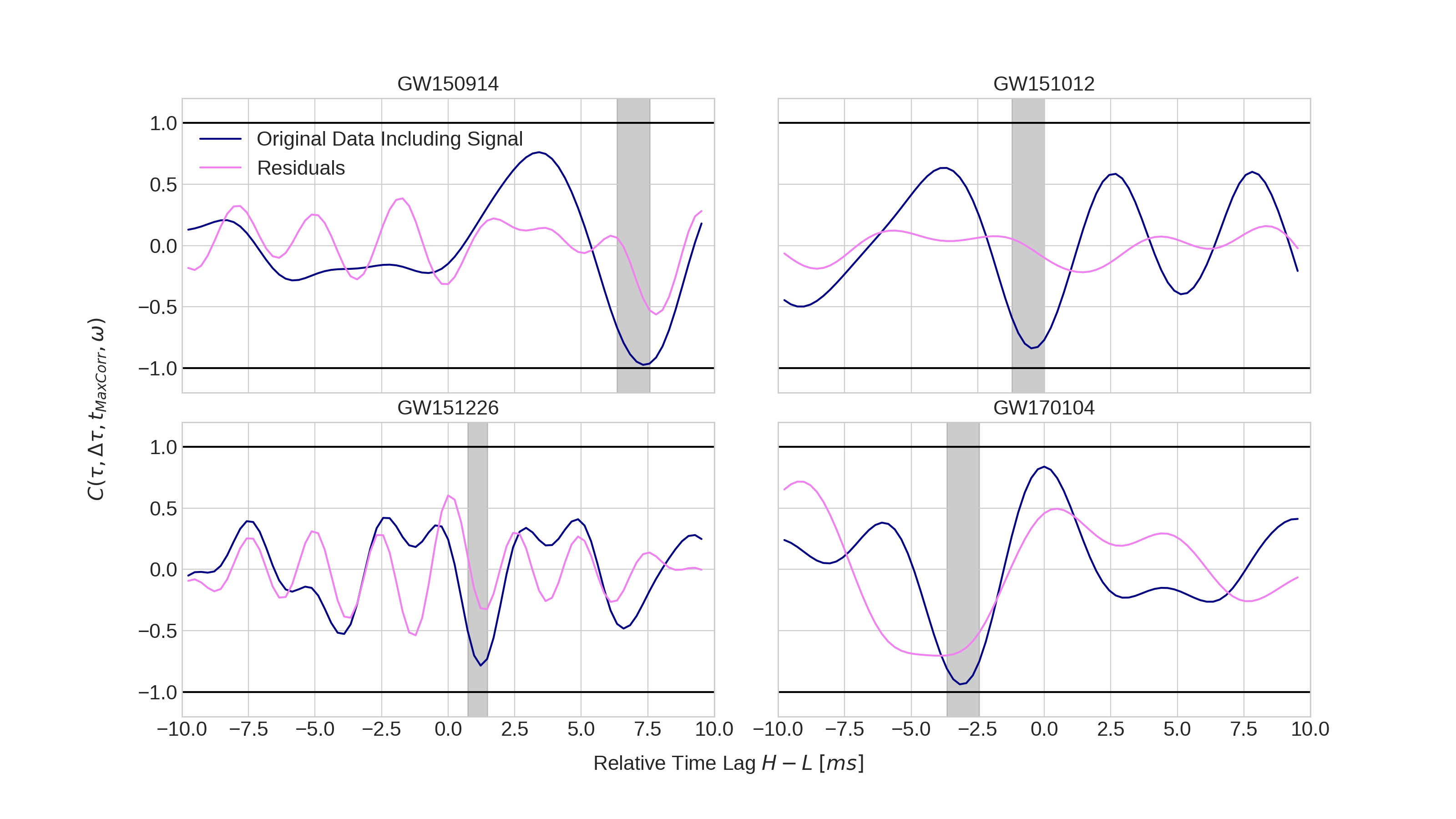}
\caption{The cross-correlations as a function of $\tau$ at the time $t=t_p$. The gray bands correspond to the strips $\tau_p\pm \Delta \tau_p$ previously used to determine $t_p$.}
\label{fig:CorrVsTimeShift}
\end{figure}

Figure~\ref{fig:CorrVsTimeShift} shows $C(\tau, \Delta \tau_p, t_p, \omega)$ as a function of $\tau$,  for $\Delta \tau_p$ and $\omega$ as in Table 
\ref{tab:table1}. The time is fixed at the values of $t_p$ obtained from the $t$ dependence analysis of $C$ and reported in Table~\ref{tab:table2}. Each panel corresponds to one of the four events we analyse. The grey band spans the time lag $\tau_p \pm \Delta \tau_p$. The figure highlights two remarkable facts.
First of all, had we determined $\tau$ and $\Delta \tau$ independently of the previous results in the literature, this would have been estimated as 
the peak of correlations and its approximate width arising in the strain data before the signal subtraction (blue line). Such peaks turn out to be within the gray bands, which proves that the values of $\tau$ used in the previous section are consistent with the data treatment of our analysis. 
Secondly, the correlation in the residuals (pink line), obtained after the subtraction of the model waveform from the original data, does not necessarily peak in the grey zone. (In some cases the global peak for the residuals is close, but not inside, the grey zone. In some cases the global peak for the residuals is a positive correlation rather than a negative correlation.)
This feature, together with the observation that the data and residual correlation peaks are well separated in Figs.~\ref{fig:150914corr}--\ref{fig:170104corr}, 
sheds light on the precision of the subtracted waveform: If there was a sizeable discrepancy between signal and subtracted waveform during the merging phase, the cross-correlation in the residuals would have likely peaked within the grey band where the signal strain amplitude is larger.\footnote{More details on this reasoning can be found in Refs.~\cite{Creswell:2017rbh, Nielsen:2018bhc}.} A peaking of the residuals' correlation within the grey band is however not sufficient for there to be a problem with the subtracted waveforms. It can also happen by pure chance.

\subsection{Background cross-correlations and statistical interpretation}
\label{sec:BkgStt}

\begin{figure}[ht]
  \centering
    \includegraphics[width=\columnwidth]{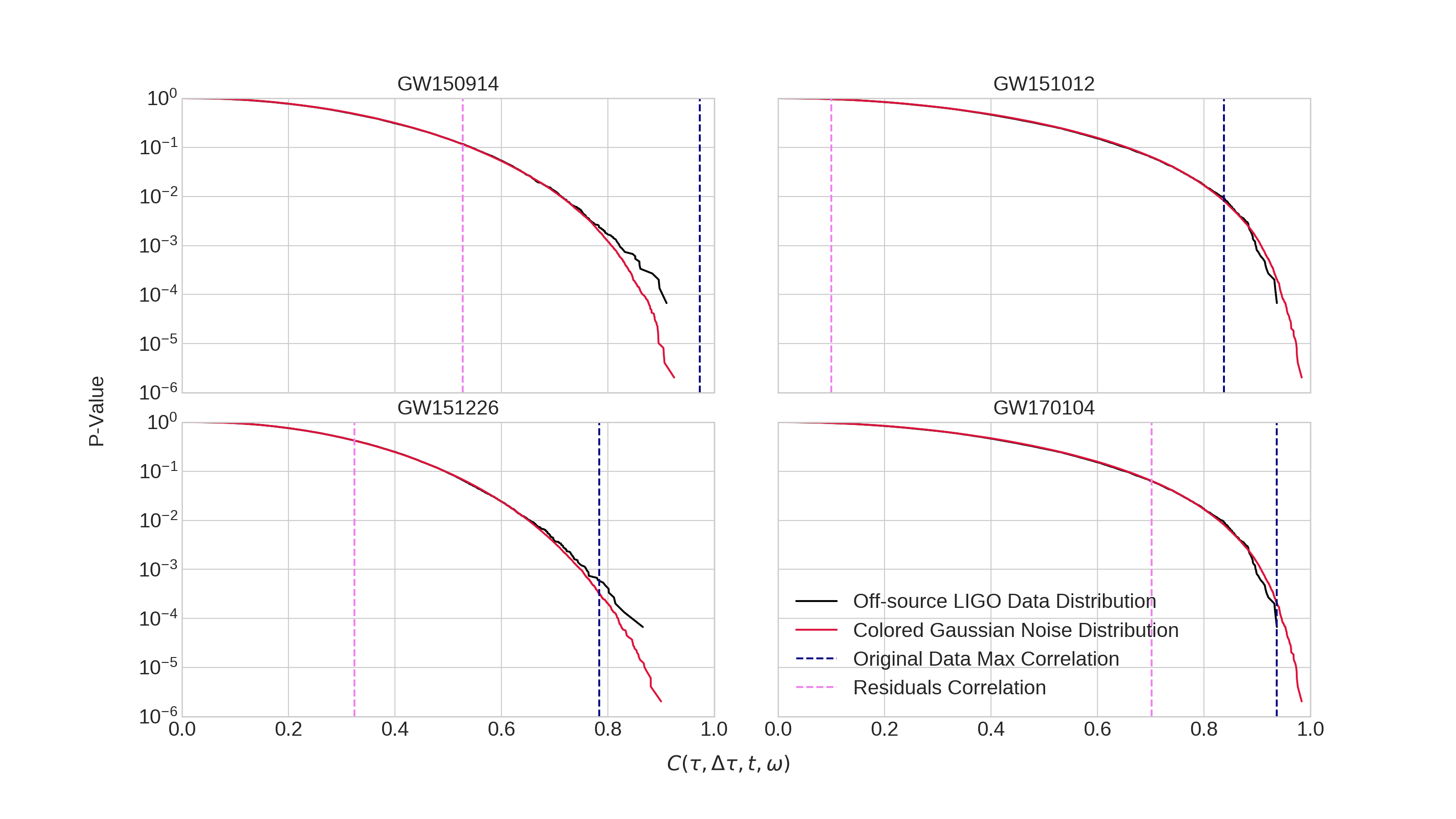}
\caption{Comparison between the actual correlation values obtained for the events, and the frequency of occurrence of similar correlation values in the background of detector time away from events (black lines) and in simulated Gaussian noise (crimson lines). Here the same background values are used for the events GW151012 and GW170104, as the same frequency bandpass range was employed for these two events.}
\label{fig:Backgrounds}
\end{figure}

In order to determine the statistical significance of our computed cross-correlations, we run the $C$ estimator over the aforementioned data sets of background, and count how often a given value of $|C|$ arises in pure noise data (preprocessed in the same way we preprocess the LIGO events that we analyze). From this we estimate the probability, in terms of a p-value, that a value at least as large as our observed cross-correlation value, would occur as a statistical fluctuation in pure noise.

Figure~\ref{fig:Backgrounds} reports the statistical significance of the cross-correlations arising in the four LIGO event data sets before and after subtracting the reconstructed signal waveform  (dashed vertical lines in blue and pink, respectively).\footnote{To reduce the computational time, the code  evaluates $R(\tau_p, t, \omega)$ instead of $C(\tau_p, \Delta \tau_p, t, \omega)$. The latter reproduces the former in the limit $\Delta \tau \to 0$ (cf.~Eqs.~\eqref{eq:corrR} and \eqref{eq:corrC}). This leads to slightly overestimate the significance of the cross-correlations $C(\tau_p, \Delta \tau_p, t_p, \omega)$ obtained in Section~\ref{sec:t-dep}.}
The black and crimson curves show the p-value deduced from  cross-correlations found in the off-source LIGO data sets and the simulated Gaussian noise data sets, respectively. 
The usage of these two different curves is complementary. The simulated Gaussian noise is guaranteed to be free of truly correlated signals while the off-source detector data more closely follows the noise distribution of the actual detectors.
The close agreement of these curves however indicates that the true noise of the detectors is statistically indistinguishable from pure Gaussian noise at the level of this test for all the bandpass frequency ranges used in our analyses.

Although this procedure of estimating significance is different in several respects from that employed by the LVC, there is still qualitative agreement in the significance ranking of the four events. From Fig.~\ref{fig:Backgrounds} the most significant event, in agreement with the LIGO results, is GW150914, with a p-value of approximately $10^{-5}$.
Moreover, the subtraction of the maximum likelihood waveform in the GW150914 data, makes $C(\tau_p, \Delta \tau_p, t, \omega)$ drop down four orders of magnitude.
The GW151226 and GW170104 data before subtraction are the next most significant events of the four analyzed, with a p-values around $10^{-3}$ and $10^{-4}$ respectively.
This similarity is unsurprising considering that both these events have a matched-filter SNR around $13$ \cite{LIGOScientific:2018mvr}.
However, after the subtraction of the numerical waveform, the p-value of the residuals' cross-correlation for GW151226 is approximately $0.2$, while for GW170104 it is around $0.08$.
This feature is a consequence of the fact that detector noise exists at all frequencies. With a bigger frequency range considered, a larger amount of noise is included in the data, making it harder to identify potential signals with a limited bandwidth.
On the other hand, once the background is bandpassed and whitened, the dependence on the frequency range of the correlations is taken into account, hence the significance obtained for the various events, can be more easily compared between events. 
Lastly, in agreement with LVC results, GW15012 is the least significant event among the four.
The p-value of its data before and after waveform subtraction is only around $10^{-2}$ and $1$, showing that the maximum likelihood waveform still efficiently reduces the cross-correlation.

In general, Fig.~\ref{fig:Backgrounds} shows that the cross-correlation method, combined with the present statistical interpretation, is a valid tool for complementary searches of signals and tests of their modelling.

\begin{table}
\centering
\begin{tabular}{| c | c | c | c |c | }
 \hline
 \textbf{Event} & \textbf{Max corr.}  &  \textbf{Max corr.~var.} &\textbf{$C(\tau_p, \Delta\tau_p,t_p, \omega)$} & \textbf{$C(\tau_p, \Delta\tau_p,t_p, \omega)$}\\
 \textbf{Name} & \textbf{time $t_p$ (s)} & \textbf{time $t_d$ (s)} &\textbf{before subtr.} & \textbf{in residuals} \\
 \hline
 GW150914 & 1126259462.39  & ~0.0069 $\pm$ \;  0.0005 &-0.97 & -0.53  \\  
 GW151012 & 1128678900.42 & -0.0006 $\pm$ \;  0.0006 &-0.84 & -0.10  \\
 GW151226 & 1135136350.62 & ~0.0011 $\pm$ \; 0.0003 &-0.78 & -0.32 \\
 GW170104 & 1167559936.57 & ~\,-0.003 $\pm$ \; 0.0005 &-0.94 & -0.70 \\
 \hline
\end{tabular}%
\caption{Time of maximum correlation and obtained correlation values at that time for the four analysed events. Correlation values are rounded up to the second decimal value and denote the maximum magnitude of correlation found in the assumed time-lag intervals, both for the original data and after subtraction of a maximum likelihood waveform. To check the full precision results see Ref.~\cite{pipeline}.
\label{tab:table2}
}
\end{table}

\section{Conclusions}
\label{sec:conclusions}
Sensitivity to the unexpected is one of the main challenges of the modern experiments. Marvelous sensitivities often come at the expense of intricate data analysis techniques required to dig out weak signals from data largely contaminated by instrumental noise. Unfortunately, such techniques tend to rely on hidden or manifest assumptions, with the risk of overlooking signatures of new physics in the data. It is then worth developing alternative techniques to analyze data with a range of different assumptions and generality. In the present paper we have investigated one of these approaches, the so called Pearson cross-correlation method. Specifically, we have employed (a variation of) it to analyse the data of the four gravitational wave events GW150914, GW151012, GW151226 and GW170104, and scrutinize some claims made about them. The study has led to the following conclusions:
\begin{itemize}
\item 
Although the Pearson cross-correlation method is less sensitive than a matched-filter analysis, it is still able to recover the events at relatively low p-values, relative to both off-source detector data and simulated Gaussian noise. 
\item No significant cross-correlation arises at about $0.1$\,s after the end of each event if the Hanford and Livingston detector data are analysed with time lags consistent with the original events. No evidence of events such as echoes is then found with this method.
\item In all of the four events, the cross-correlations in the data after subtracting maximum likelihood general-relativity waveforms, is consistent with noise fluctuations. Specifically, such cross-correlations have p-values larger than $0.05$. This is consistent with the residual analysis results of the LVC~\cite{LIGOScientific:2019fpa}, and amply compatible with general relativity. 
\item The maximum peaks in the residual correlations do not typically occur at detector time lags consistent with the original events. This seems statistically natural in the case that the subtracted waveforms well fit the signal in the merging stage, which is the phase at which the binary signal strain is the largest and the Pearson cross-correlation method consequently reaches its highest sensitivity.
\end{itemize}

Several aspects however remain to be explored in more depth. With more computational resources dedicated to the study, it would be interesting to systematically investigate the performances of the Pearson cross-correlation method in the whole catalogue of LIGO events, and excise these data with a multivariational approach of the Pearson cross-correlation estimator (here we have run it by varying either the GPS detector time or the time lag). This is especially true of much lower mass systems such as the binary neutron star merger GW170817. While analysis of this event is beyond the scope of the present work, we see no fundamental reasons why the current techniques could not be applied there. The same method could also be adopted to test the quality of the multi-source fitting necessary to solve the ``enchilada'' problem in LISA data~\cite{Babak:2017cjl}. We plan to deal with some of these aspects in future investigations.

\acknowledgments
This work uses open-source code presented in Ref.~\cite{Nielsen:2018bhc}. AN and FF thank the Max Planck Institute for Gravitational Physics in Hanover, Germany for support and hospitality during the initial stages of this project. G.N.~is partly supported by the Research Council of Norway, ROMFORSK grant, project.~no.~302640.

\appendix

\section{Maximum likelihood IMR waveform parameters}

 In this appendix, we report the values of the general relativity model templates that we subtract from the detectors data for the four analyzed events. These values correspond to the maximum likelihood values reported in ref.~\cite{Biwer:2018osg}.
 They are constructed using the phenomenological inspiral-merger-ringdown waveform family IMRPhenomPv2 \cite{Hannam:2013oca} which is freely available as part of LALSuite \cite{LALsuite}.

\begin{table}
\centering
\resizebox{\columnwidth}{!}{
\begin{tabular}{| c | c | c | c | c | c | c | }
 \hline
 \textbf{Parameter} & \textbf{Description}  &  \textbf{GW150914} & \textbf{GW151012} & \textbf{GW151226} & \textbf{GW170104}\\
 \hline
 $m_1$ & Mass of the larger black hole $(M_{\bigodot})$ & $39$ & $23$ & $19$ & $39$\\
 $m_2$ & Mass of the smaller black hole $(M_{\bigodot})$ & $32$ & $19$ & $7$ & $21$\\
 $a_1$ & Dimensionless spin of the larger BH & $0.977$ & $0.299$ & $0.607$ & $0.550$\\
 $\theta_1^a$ & Azth. angle of the larger BH spin $(rad)$ & $3.6$ & $1.0$ & $2.5$ & $3.49$\\
 $\theta_1^p$ & Polar angle of the larger BH spin $(rad)$ & $1.6$ & $2.30$ & $1.2$ & $2.39$\\
 $a_2$ & Dimensionless spin of the smaller BH & $0.189$ & $0.067$ & $0.938$ & $0.553$\\
 $\theta_2^a$ & Azth. angle of the smaller BH spin $(rad)$  & $3.44$ & $5.48$ & $5.32$ & $0.06$\\
 $\theta_2^p$ & Polar angle of the smaller BH spin $(rad)$  & $2.49$ & $0.40$ & $1.05$ & $0.59$\\
 $d_L$ & Luminosity distance $(MPc)$ & $480$ & $750$ & $380$ & $530$\\
 $\alpha$ & Right ascension $(rad)$ & $1.57$ & $0.65$ & $1.85$ & $0.89$\\
 $\delta$ & Declination $(rad)$ & $-1.27$ & $0.07$ & $0.99$ & $-0.80$\\
 $\psi$ & Polarization $(rad)$ & $5.99 $ & $5.64$ & $2.76$ & $5.69$\\
 $f_0$ & Starting frequency of the waveform $(Hz)$  & $10$ & $10$  & $10$ & $10$\\
 $f_{ref}$ & Reference frequency $(Hz)$ & $20$ & $20$ & $20$ & $20$\\
 $i$ & Inclination of the binary at $f_{ref} \; (rad)$ & $2.91$ & $2.32$ & $0.66$ & $1.09$\\
 $\phi_c$ & Reference phase at $f_{ref}$ & $0.69$ & $4.44$ & $ \emptyset $ & $\emptyset$\\
 $\Delta \phi$ & Waveform’s phase with respect to $\phi_c$ & $-0.92$ & $-0.91$ & $-0.10$ & $-1.80$\\
 \hline
\end{tabular}%
}
\caption{Table of the maximum likelihood parameters for the waveforms of the four events considered here. Values are rounded up to an arbitrary precision. The value of $\phi_c$ for GW151226 and GW170104 is not reported as it is not part of the maximum likelihood parameters reported by \cite{Biwer:2018osg}. For the full precision results see Ref.~\cite{pipeline}.
\label{tab:table3}
}
\end{table}

\bibliography{bibfile.bib} 

\begin{thebibliography}{10}

\bibitem{Abbott:2016blz}
B.~Abbott {\em et~al.}, ``{Observation of Gravitational Waves from a Binary
  Black Hole Merger},'' {\em Phys. Rev. Lett.}, vol.~116, no.~6, p.~061102,
  2016.

\bibitem{TheLIGOScientific:2016agk}
B.~Abbott {\em et~al.}, ``{GW150914: The Advanced LIGO Detectors in the Era of
  First Discoveries},'' {\em Phys. Rev. Lett.}, vol.~116, no.~13, p.~131103,
  2016.

\bibitem{Brugmann366}
B.~Br{\"u}gmann, ``Fundamentals of numerical relativity for gravitational wave
  sources,'' {\em Science}, vol.~361, no.~6400, pp.~366--371, 2018.

\bibitem{Allen:2005fk}
B.~Allen, W.~G. Anderson, P.~R. Brady, D.~A. Brown, and J.~D. Creighton,
  ``{FINDCHIRP: An Algorithm for detection of gravitational waves from
  inspiraling compact binaries},'' {\em Phys. Rev. D}, vol.~85, p.~122006,
  2012.

\bibitem{Usman:2015kfa}
S.~A. Usman {\em et~al.}, ``{The PyCBC search for gravitational waves from
  compact binary coalescence},'' {\em Class. Quant. Grav.}, vol.~33, no.~21,
  p.~215004, 2016.

\bibitem{Sachdev:2019vvd}
S.~Sachdev {\em et~al.}, ``{The GstLAL Search Analysis Methods for Compact
  Binary Mergers in Advanced LIGO's Second and Advanced Virgo's First Observing
  Runs},'' 1 2019.

\bibitem{Klimenko:2015ypf}
S.~Klimenko {\em et~al.}, ``{Method for detection and reconstruction of
  gravitational wave transients with networks of advanced detectors},'' {\em
  Phys. Rev. D}, vol.~93, no.~4, p.~042004, 2016.

\bibitem{Lynch:2015yin}
R.~Lynch, S.~Vitale, R.~Essick, E.~Katsavounidis, and F.~Robinet,
  ``{Information-theoretic approach to the gravitational-wave burst detection
  problem},'' {\em Phys. Rev. D}, vol.~95, no.~10, p.~104046, 2017.

\bibitem{LIGOScientific:2019fpa}
B.~Abbott {\em et~al.}, ``{Tests of General Relativity with the Binary Black
  Hole Signals from the LIGO-Virgo Catalog GWTC-1},'' {\em Phys. Rev. D},
  vol.~100, no.~10, p.~104036, 2019.

\bibitem{Salemi:2019uea}
F.~Salemi, E.~Milotti, G.~Prodi, G.~Vedovato, C.~Lazzaro, S.~Tiwari,
  S.~Vinciguerra, M.~Drago, and S.~Klimenko, ``{Wider look at the
  gravitational-wave transients from GWTC-1 using an unmodeled reconstruction
  method},'' {\em Phys. Rev. D}, vol.~100, no.~4, p.~042003, 2019.

\bibitem{Tsang:2019zra}
K.~W. Tsang, A.~Ghosh, A.~Samajdar, K.~Chatziioannou, S.~Mastrogiovanni,
  M.~Agathos, and C.~Van Den~Broeck, ``{A morphology-independent search for
  gravitational wave echoes in data from the first and second observing runs of
  Advanced LIGO and Advanced Virgo},'' {\em Phys. Rev. D}, vol.~101, no.~6,
  p.~064012, 2020.

\bibitem{Edelman:2020aqj}
B.~Edelman {\em et~al.}, ``{Constraining Unmodeled Physics with Compact Binary
  Mergers from GWTC-1},'' {\em arXiv:2008.06436}.

\bibitem{Monitor:2017mdv}
B.~Abbott {\em et~al.}, ``{Gravitational Waves and Gamma-rays from a Binary
  Neutron Star Merger: GW170817 and GRB 170817A},'' {\em Astrophys. J. Lett.},
  vol.~848, no.~2, p.~L13, 2017.

\bibitem{Abbott:2020uma}
B.~Abbott {\em et~al.}, ``{GW190425: Observation of a Compact Binary
  Coalescence with Total Mass $\sim 3.4 M_{\odot}$},'' {\em Astrophys. J.
  Lett.}, vol.~892, no.~1, p.~L3, 2020.

\bibitem{Barausse:2014tra}
E.~Barausse, V.~Cardoso, and P.~Pani, ``{Can environmental effects spoil
  precision gravitational-wave astrophysics?},'' {\em Phys. Rev. D}, vol.~89,
  no.~10, p.~104059, 2014.

\bibitem{Liu:2016kib}
H.~Liu and A.~D. Jackson, ``{Possible associated signal with GW150914 in the
  LIGO data},'' {\em JCAP}, vol.~1610, no.~10, p.~014, 2016.

\bibitem{Creswell:2017rbh}
J.~Creswell, S.~von Hausegger, A.~D. Jackson, H.~Liu, and P.~Naselsky, ``{On
  the time lags of the LIGO signals},'' {\em JCAP}, vol.~08, p.~013, 2017.

\bibitem{Liu:2018dgm}
H.~Liu, J.~Creswell, S.~von Hausegger, A.~D. Jackson, and P.~Naselsky, ``{A
  blind search for a common signal in gravitational wave detectors},'' {\em
  JCAP}, vol.~02, p.~013, 2018.

\bibitem{Nielsen:2018bhc}
A.~B. Nielsen, A.~H. Nitz, C.~D. Capano, and D.~A. Brown, ``{Investigating the
  noise residuals around the gravitational wave event GW150914},'' {\em JCAP},
  vol.~02, p.~019, 2019.

\bibitem{Jackson:2019xbq}
A.~D. Jackson, H.~Liu, and P.~Naselsky, ``{Noise residuals for GW150914 using
  maximum likelihood and numerical relativity templates},'' {\em JCAP},
  vol.~1905, p.~014, 2019.

\bibitem{Maroju:2019ymy}
R.~Maroju, S.~R. Dyuthi, A.~Sukrutha, and S.~Desai, ``{Looking for ancillary
  signals around GW150914},'' {\em JCAP}, vol.~04, p.~007, 2019.

\bibitem{LIGOScientific:2019hgc}
B.~P. Abbott {\em et~al.}, ``{A guide to LIGO--Virgo detector noise and
  extraction of transient gravitational-wave signals},'' {\em Class. Quant.
  Grav.}, vol.~37, no.~5, p.~055002, 2020.

\bibitem{Nitz:2018imz}
A.~H. Nitz, C.~Capano, A.~B. Nielsen, S.~Reyes, R.~White, D.~A. Brown, and
  B.~Krishnan, ``{1-OGC: The first open gravitational-wave catalog of binary
  mergers from analysis of public Advanced LIGO data},'' {\em Astrophys. J.},
  vol.~872, no.~2, p.~195, 2019.

\bibitem{LIGOScientific:2018mvr}
B.~Abbott {\em et~al.}, ``{GWTC-1: A Gravitational-Wave Transient Catalog of
  Compact Binary Mergers Observed by LIGO and Virgo during the First and Second
  Observing Runs},'' {\em Phys. Rev. X}, vol.~9, no.~3, p.~031040, 2019.

\bibitem{Venumadhav:2019tad}
T.~Venumadhav, B.~Zackay, J.~Roulet, L.~Dai, and M.~Zaldarriaga, ``{New search
  pipeline for compact binary mergers: Results for binary black holes in the
  first observing run of Advanced LIGO},'' {\em Phys. Rev. D}, vol.~100, no.~2,
  p.~023011, 2019.

\bibitem{Vallisneri:2014vxa}
M.~Vallisneri, J.~Kanner, R.~Williams, A.~Weinstein, and B.~Stephens, ``{The
  LIGO Open Science Center},'' {\em J. Phys. Conf. Ser.}, vol.~610, no.~1,
  p.~012021, 2015.

\bibitem{De:2018zrk}
S.~De, C.~M. Biwer, C.~D. Capano, A.~H. Nitz, and D.~A. Brown, ``{Posterior
  samples of the parameters of binary black holes from Advanced LIGO, Virgo's
  second observing run},'' 11 2018.

\bibitem{Biwer:2018osg}
C.~Biwer, C.~D. Capano, S.~De, M.~Cabero, D.~A. Brown, A.~H. Nitz, and
  V.~Raymond, ``{PyCBC Inference: A Python-based parameter estimation toolkit
  for compact binary coalescence signals},'' {\em Publ. Astron. Soc. Pac.},
  vol.~131, no.~996, p.~024503, 2019.

\bibitem{TheLIGOScientific:2016pea}
B.~Abbott {\em et~al.}, ``{Binary Black Hole Mergers in the first Advanced LIGO
  Observing Run},'' {\em Phys. Rev. X}, vol.~6, no.~4, p.~041015, 2016.
\newblock [Erratum: Phys.Rev.X 8, 039903 (2018)].

\bibitem{pipeline}
P.~Marcoccia.
\newblock https://github.com/GravWaves-IMF/Correlation-Method-first-2019.

\bibitem{Babak:2017cjl}
S.~Babak, ``{``Enchilada'' is back on the menu},'' {\em J. Phys. Conf. Ser.},
  vol.~840, no.~1, p.~012026, 2017.

\bibitem{Hannam:2013oca}
M.~Hannam, P.~Schmidt, A.~Boh\'e, L.~Haegel, S.~Husa, F.~Ohme, G.~Pratten, and
  M.~P\"urrer, ``{Simple Model of Complete Precessing Black-Hole-Binary
  Gravitational Waveforms},'' {\em Phys. Rev. Lett.}, vol.~113, no.~15,
  p.~151101, 2014.

\bibitem{LALsuite}
R.~A. Mercer {\em et~al.}, ``Ligo algorithm library v6.49.,'' 2018.
\newblock https://git.ligo.org/lscsoft/lalsuite.

\end{thebibliography}
\bibliographystyle{ieeetr}

\end{document}